\documentclass[preprint,12pt]{elsarticle}
\usepackage{amssymb}
\usepackage{graphicx}

\usepackage{lineno}

\begin{document}

\begin{frontmatter}
  
\title{Comparisons of the MINOS Near and Far Detector Readout Systems
  at a Test Beam}
  
  \author[ox]{A.~Cabrera\fnref{apc}\corref{cor1}}\ead{anatael@in2p3.fr}
  \author[ucl]{P.~Adamson}
  \author[ox]{M.~Barker}
  \author[ral]{A.~Belias}
  \author[pitt]{S.~Boyd}
  \author[ucl]{G.~Crone}
  \author[arg]{G.~Drake}
  \author[susx]{E.~Falk}
  \author[susx]{P.G.~Harris}
  \author[ox,ral]{J.~Hartnell\corref{cor1}}\ead{j.j.hartnell@sussex.ac.uk}
  \author[ucl]{L.~Jenner}
  \author[ut]{M.~Kordosky}
  \author[ut]{K.~Lang}
  \author[ox]{R.P.~Litchfield}
  \author[cal]{D.~Michael\fnref{de}}
  \author[ox]{P.S.~Miyagawa}
  \author[susx]{R.~Morse}
  \author[st]{S.~Murgia}
  \author[ucl]{R.~Nichol}
  \author[ral]{T.~Nicholls}
  \author[ral]{G.F.~Pearce}
  \author[umn]{D.~Petyt}
  \author[arg]{D.~Reyna}
  \author[ucl]{R.~Saakyan}
  \author[fnal]{P.~Shanahan}
  \author[ucl]{C.~Smith}
  \author[susx]{P.~Symes}
  \author[ox]{N.~Tagg\fnref{ott}}
  \author[ucl]{J.~Thomas}
  \author[ut]{P.~Vahle}
  \author[ox]{A.~Weber}

  \address[arg]{Argonne National Laboratory, Argonne, Illinois 60439, USA}
  \address[cal]{Lauritsen Laboratory, California Institute of Technology, Pasadena, California 91125, USA}
  \address[fnal]{Fermi National Accelerator Laboratory, Batavia, Illinois 60510, USA}
  \address[ucl]{Department of Physics and Astronomy, University College London, Gower Street, London WC1E 6BT, UK}
  \address[umn]{University of Minnesota, Minneapolis, Minnesota 55455, USA}
  \address[ox]{Subdepartment of Particle Physics, University of Oxford, Denys Wilkinson Building, Keble Road, Oxford OX1 3RH, UK}
  \address[pitt]{Department of Physics and Astronomy, University of Pittsburgh, Pittsburgh, Pennsylvania 15260, USA}
  \address[ral]{Rutherford Appleton Laboratory, Chilton, Didcot, Oxfordshire, OX11 OQX, UK}
  \address[st]{Department of Physics, Stanford University, Stanford, California 94305, USA}
  \address[susx]{Department of Physics and Astronomy, University of Sussex, Falmer, Brighton BN1 9QH, UK}
  \address[ut]{Department of Physics, University of Texas at Austin, Austin, Texas 78712, USA}
  \fntext[apc]{Now at APC, 10 rue Alice Domon et L\'{e}onie Duquet, F-75205 Paris Cedex 13, France}
  \fntext[ott]{Now at Department of Physics and Astronomy, Otterbein College, 1 Otterbein College, Westerville, OH 43081-2006, USA}
  \cortext[cor1]{Corresponding authors, Tel.: +33\,1\,57\,27\,61\,64; fax: +33\,1\,57\,27\,60\,71}
  \fntext[de]{Deceased}


\begin{abstract}
  MINOS is a long baseline neutrino oscillation experiment that uses
  two detectors separated by 734~km. The readout systems used for the
  two detectors are different and have to be independently
  calibrated. To verify and make a direct comparison of the calibrated
  response of the two readout systems, test beam data were acquired
  using a smaller calibration detector. This detector was
  simultaneously instrumented with both readout systems and exposed to
  the CERN PS T7 test beam. Differences in the calibrated response of
  the two systems are shown to arise from differences in response
  non-linearity, photomultiplier tube crosstalk, and threshold effects
  at the few percent level. These differences are reproduced by the
  Monte Carlo (MC) simulation to better than 1\% and a scheme that
  corrects for these differences by calibrating the MC to match the
  data in each detector separately is presented. The overall
  difference in calorimetric response between the two readout systems
  is shown to be consistent with zero to a precision of 1.3\% in data
  and 0.3\% in MC with no significant energy dependence.
\end{abstract}



\end{frontmatter}



\section{Introduction}

The Main Injector Neutrino Oscillation Search (MINOS) is a long
baseline, two-detector neutrino oscillation experiment that uses the
NuMI neutrino beam at Fermilab~\cite{ref:NuMI1}. The energy
spectrum and flavour composition of the neutrino beam is measured at
two detectors located 734~km apart: the Near detector (ND) at Fermilab
and the Far detector (FD) at the Soudan Underground Laboratory in
Minnesota~\cite{ref:MINOSDETNIM}. MINOS is sensitive to neutrino
oscillations in the region studied by the atmospheric neutrino
experiments and has recently measured $|\Delta m^2|=2.43 \pm 0.13
\times 10^{-3}~\rm eV^2$~\cite{ref:MINOSCCPRL2008}.

To control systematic errors, such as those that arise from
uncertainties in neutrino flux and cross-sections, MINOS was designed
as a two-detector experiment. This design allows a relative
measurement to be made but it subsequently becomes necessary to ensure
a precise inter-detector calibration of the energy scale. The MINOS ND
and FD have a similar steel-scintillator structure but were
instrumented with different readout systems. The driving factors for
using different readout systems were the relatively small size and
high event rate of the ND compared to the large size and low event
rate of the FD. To investigate the effects of the two readout systems
on the relative energy scale, test beam data were taken with the
dedicated MINOS Calibration detector (CalDet)~\cite{ref:CalDet1} that
was simultaneously instrumented with both readout
systems. Specifically, differences arising from response
non-linearity, photomultiplier tube crosstalk, and threshold effects
were studied.

This paper quantifies the differences between the two readout systems
and demonstrates how they can be controlled. In
Sections~\ref{sec:detdet} and~\ref{sec:selection} the experimental setup
and the selection of the events used in the analysis are described.
Section~\ref{sec:calib} outlines the calibration scheme and
Section~\ref{sec:calibsyst} details the uncertainties on the
calibration. Comparisons of strip occupancy and PMT crosstalk for the
two readout systems are given in Section~\ref{sec:stripOccAndPmtXTalk}.
The results of the calorimetric response comparisons and the conclusions
are given in Section~\ref{sec:results} and Section~\ref{sec:conclusions}
respectively. Further details of this analysis can be found
in~\cite{ref:MyThesis}.


\section{The MINOS Calibration Detector}
\label{sec:detdet}

The MINOS detectors are tracking-sampling calorimeters consisting of
alternating layers of steel and scintillator planes. CalDet planes use
2.50~cm thick unmagnetised steel in contrast with the ND and FD that
use 2.54~cm thick magnetised steel. In all three MINOS detectors, the
plastic scintillator planes are made out of strips which are 1~cm
thick and 4.1~cm wide. Successive scintillator planes are rotated by
90$^\circ$ to allow three dimensional event reconstruction. CalDet
consists of 60 steel planes measuring 1~$\times$~1~m square,
interleaved with scintillator planes comprising of 24 strips of 1~m
length. Light produced in the scintillator is captured by 1.2~mm
diameter wavelength shifting (WLS) fibre optic cables. At two corners
of each CalDet plane the WLS fibres from the 24 strips are brought
together in a manifold. Additional fibre optic cables are then
attached at the end of the manifolds to guide the light to multi-anode
photomultiplier tubes (PMTs). The light output of CalDet was
engineered to match the ND and FD by using long fibres: 6~m clear
fibres on one end of the strips and 3~m green WLS
fibres\footnote{Previously 4~m green fibres were used: these fibres
  were converted into a set of 1~m and 3~m fibres in 2003.} on the
opposite end.

\begin{figure}
  \centerline{\includegraphics[clip,width=\linewidth]{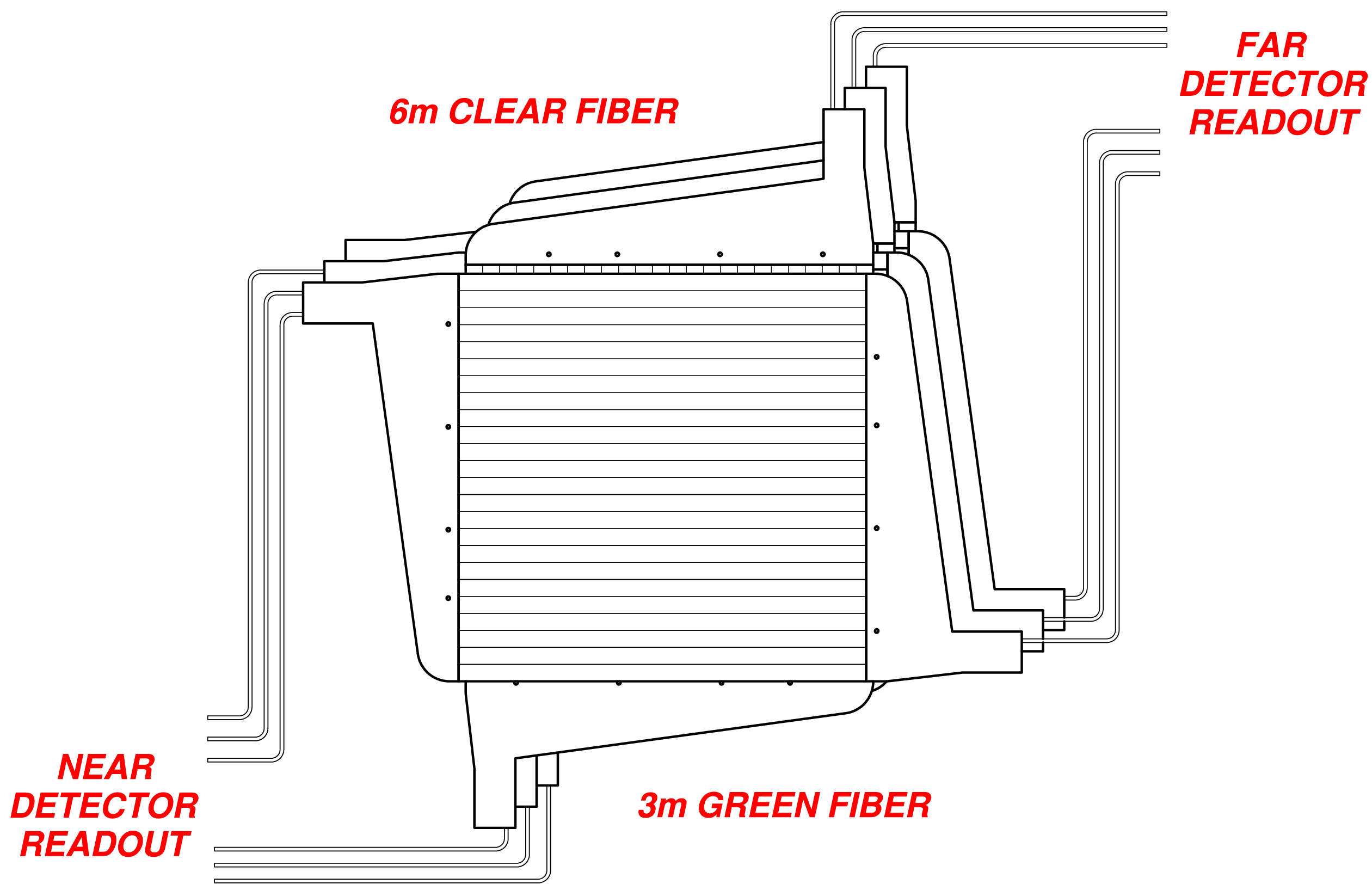}}
\caption{A schematic diagram showing the first 6 planes of CalDet.  In
  the configuration shown, CalDet was simultaneously instrumented with
  both ND and FD readout systems. Clear fibres are connected to the
  two upper manifolds and green WLS fibres to the two lower
  manifolds.}
\label{fig:det}
\end{figure}

A schematic diagram showing the configuration of CalDet when it was
simultaneously instrumented with both readout systems is shown in
Fig.~\ref{fig:det}.  The view of the detector shown is looking
downstream at the first plane with successive planes following into
the page. Each scintillator strip was read out at both ends: one end
was instrumented with the ND readout system while the opposite end
used the FD readout system. Successive planes alternate between the ND
readout system being connected via clear fibres and the FD readout
system by green WLS fibres and vice versa.

The ND uses 64-anode Hamamatsu M64 PMTs~\cite{ref:M64} with a custom
base divider ratio of 3-2-2-1-1-1-1-1-1-1-1-2-5 and electronics based
on the QIE ASIC~\cite{ref:QIE}. The FD readout system uses 16-anode
Hamamatsu M16 PMTs~\cite{ref:M16} with a custom base divider ratio of
2.4-2.4-2.4-1-1-1-1-1-1-1-1-1.2-2.4 and electronics based on the VA
ASIC~\cite{ref:VAFEE1}.

The multi-anode PMTs used in MINOS have a single photocathode. The
region of the cathode from which charge is focused on to a specific
anode is called a pixel. The PMTs were housed in custom made boxes
with optical routing that guided the light from individual strips to
specific pixels. The routing pattern was designed such that
neighbouring strips did not use neighbouring pixels. This was done to
minimise the effect of crosstalk on pattern recognition. Crosstalk
predominantly occurs on pixels next to the one illuminated due to
leakage of photoelectrons before the first
dynode~\cite{ref:M64,ref:M16}. The optical routing of the FD readout
system was such that three PMTs were fully utilised to readout two
planes. The ND PMTs have 64 pixels, hence only one PMT was required to
read out two detector planes and 16 pixels remained unilluminated.

In the ND front-end electronics, the charge from each PMT pixel is
integrated and digitised continuously at 53 MHz (every 18.8 ns) for up
to several tens of $\rm \mu$s with no dead-time. The digitisation is
performed by a specialised ASIC, the QIE-chip. It splits the incoming
signal into eight binary-weighted ``ranges'' and integrates the
resulting fractional currents on capacitors in each range. A
sample-and-hold circuit stores the integrated charge in each
range. The QIE-chip then selects the first non-saturated range and
passes its held voltage to the 8-bit ADC for digitisation. This
combination of the 8 ranges with the 8-bit ADC gives the QIE-chip the
same dynamic range as a 16-bit ADC but with an approximately constant
fractional error across the entire dynamic range. The sensitivity of
the ND front-end electronics is 1.4~fC/ADC\@.

The QIE circuit does not provide a linear response by itself because
each range has a different ADC offset and gain. A 16-bit DC current
injection circuit is used to map the response of the chip. The results
of that calibration are uploaded locally in a look-up-table and used
during normal data taking. The calibrated output of the ND electronics
is linear over the entire dynamic range to better than 0.5\%. The
stored data are processed once the acquisition period (of up to
several tens of $\rm \mu$s) is complete: pedestal subtraction and
linearisation are performed for each channel via the look-up-table in
a single step.

The FD readout system is based on the front-end ASIC VA32\_HDR11
(short VA chip), developed in collaboration with the Norwegian company
IDEA~ASA\@.  The chip includes a charge sensitive preamplifier, a
shaper, sample and hold for each channel, and is followed by an
analogue output multiplexer. The output from a selected channel can be
switched to a differential output buffer, which drives a significant
length of cable to a remote ADC\@. The ADC value is proportional to
the integrated charge of the PMT pulse with a sensitivity of 2~fC/ADC,
but saturation starts to occur between 15\,-\,30~pC. Readout of the
front-end electronics is triggered by a signal from the last dynode of
each PMT and the dead time after each digitisation is 6~$\rm
\mu$s. This electronics was a cost effective choice for the low rate
environment of a deep underground detector.

The amount of recorded data is reduced in both systems by only
recording hits above a readout threshold, which is based on the
electronics pedestal width. The ND electronics has a 20~ADC threshold
for every channel and this is applied to each 18.8~ns integration
sample. For the FD electronics a different threshold is applied
separately to the total integrated charge on each channel; the average
threshold is 18~ADCs with a rms spread of 3~ADCs. FD electronics
threshold calculation was performed at least daily and the ND
electronics was calibrated just once. Both sets of electronics
operated stably with no significant threshold variation over the
duration of the data taking period (many weeks).

\subsection{The PS Test Beam and External Trigger}
\label{sec:detbeam}

The results presented in this paper are based on data obtained by
exposing CalDet in the T7 test beam in the East Experimental Hall of
the 24~GeV/c CERN Proton Synchrotron (PS)~\cite{ref:T7}. The dual
polarity, mixed composition beam of e, $\rm \mu$, $\rm \pi$ and p was
operated between 0.6\,-\,10~GeV/c. The beam was measured to have an
rms width of 2~cm in both the horizontal and vertical projections
using CalDet. The beam line was instrumented with several Cherenkov
counters filled with CO$_2$ to identify electrons and a time-of-flight
(TOF) system to further aid particle identification. The Cherenkov
counters had a combined efficiency of 99.9$\pm0.1$\% below 3~GeV/c and
96$\pm1$\% between 3\,-\,6~GeV/c~\cite{ref:Mike}. The TOF system had a
timing resolution of between 110\,-\,120~ps and a baseline of
9.1~m. Control of the beam momenta spread ($\Delta p/p <2\%$) and the
instantaneous event rate ($<1$~kHz) was achieved by adjusting the
brass beam line collimators. For further details
see~\cite{ref:CalDet1,ref:Mike}.

Synchronous readout of both systems was accomplished by an external
trigger provided by the TOF counters. After a trigger, the ND and FD
electronics were enabled for 376~ns and 1~$\mathrm \mu$s time windows
respectively, which was long enough to record individual events. In
addition, for 50~$\rm \mu$s after each trigger, signals from the TOF
system were suppressed to ensure that the electronics had ample time
to complete the digitisation of the first event.

\subsection{The Monte Carlo Simulation}
\label{sec:detmc}

The simulation results presented alongside the data in this paper use
the MINOS Monte Carlo (MC) simulation. The MC is based on
\textsl{GEANT3}~\cite{ref:geant} and is used to generate raw energy
depositions (\textsl{GEANT} hits), which serve as the input to a C++
based detector response model. The response model includes the effects
of light collection and propagation from the scintillator strips to
the PMT photocathode as well as the PMT and the electronics.

The attenuation of light along the length of a strip is modelled as
the sum of two exponential functions to give a short and a long
attenuation component. Fits to both bench-top and cosmic ray data
determined the lengths and relative contributions of the long and
short components of the attenuation. In the simulation of the CalDet
the approximation is made that every strip has the same short and long
attenuation components. The variation in the light output of the
strips and the gains of the PMTs are directly incorporated into the
simulation by using the calibration constants determined by the
detector calibration procedure (see Section~\ref{sec:calib}).


\section{Event Selection}
\label{sec:selection}

The readout system comparison required separation of specific particle
types and single-particle interactions in CalDet. Cuts to achieve this
were developed and are presented here. Positron and muon samples were
chosen because they are efficiently identified with high purity and
their energy loss mechanisms are well understood. When passing through
iron at GeV-scale energies the positrons form a dense electromagnetic
shower, which makes them ideal candidates for probing the full range
of energy depositions expected for neutrino interactions in
MINOS\@. The muons were used to probe some of the systematic errors of
the calibration, their energy loss was close to minimum ionising and
considerably more uniform through the detector than positrons.

Strict timing cuts were made to ensure that only events with single
particle interactions were selected, and that events with multiple
overlapping events from the test beam were rejected. This was done
since rate of events at the FD is so low (about 1~Hz of cosmic ray
muons) there is little to be gained from comparing the response for
multiple overlapping particle interactions occurring at the same time
(``pile-up'' events). These timing cuts removed between 10\,-\,25\% of
events depending on the beam conditions. The events removed were
studied and had timing distributions with distinct peaks, consistent
with more than one particle interaction. After these timing cuts: the
Cherenkov counter registered a hit consistent with two simultaneous
positrons in 1\% of events; and a 20 plane cut on event length removed
less than 0.1\% of events in the positron sample. After applying all
the cuts, a cross check based on the energy of the event was
performed: less than 0.1\% (0.5\%) of the selected events had an
energy consistent with two simultaneous beam particles in the positron
(muon) samples. Pile-up at this level had a completely negligible
affect on the analysis. In addition, further timing cuts were made to
ensure that the single particle interactions were measured fully by
both readout systems, since the length of time the two systems were
enabled was different.

Positrons were selected by demanding the following: a TOF consistent
with their velocity; a Cherenkov counter hit compatible with a single
positron; coincidence in time of the TOF trigger and Cherenkov
counter; and energy depositions only in the first 20 planes (to
further reduce the number of pile-up events, specifically the case
where an positron arrives in coincidence with a longer muon or
pion). Contamination of the positron sample from pions that caused a
Cherenkov counter hit was less than 1\% across all
energies~\cite{ref:Trish} and had a negligible effect on the analysis.

High energy muons were selected by demanding: a TOF consistent with
their velocity; no signal in the Cherenkov counter; and a particle
track crossing all 60 planes of the detector. The Cherenkov counter
efficiencies coupled with the falling beam fraction of positrons at
higher energies reduced the contamination from positrons in the
muon/pion sample to less than 0.5\% at entry to the detector. The
60~plane requirement eliminated any remaining positron events and
reduced the pion contamination to less than 1\%.


\section{Calibration}
\label{sec:calib}

The intrinsic calorimetric response of the ND and FD readout systems is
different and thus it is necessary to calibrate their response
independently. The readout system comparisons presented in this paper
are dependent on the calibration and hence constitute a precise test of
the calibration procedure's ability to remove the intrinsic response
differences between the two readout systems.

The calibration of CalDet relies on an LED-based light injection
system~\cite{ref:LI} and cosmic ray muons. A detailed description of
the calibration techniques developed for MINOS using CalDet are given
in~\cite{ref:CalDet1}. A multi-stage procedure that converts the raw
charge $Q_{raw}(i,t,x)$ measured by strip $i$ at time $t$ and distance
from the centre of the strip $x$ into a corrected signal $Q_{cor}$ was
used. Each calibration stage produced a numerical factor
(``calibration constant''). $Q_{cor}$ is defined by the product of
$Q_{raw}(i,t,x)$ and the calibration constant from each stage:
\begin{eqnarray}
\label{eq:calib}
Q_{cor} = Q_{raw}(i,t,x) \times D(i,t) \nonumber\\ 
         \times U(i) \times A(i,x) \times S,\nonumber
\end{eqnarray}
where $D$, $U$, $A$ and $S$ refer to:

{\bf Drift Correction $D(i,t)$:} To correct for temporal variations,
data from the light injection system were used to track the response
of each channel in the two readout systems\footnote{The light
  injection data were also used to determine the gain of each channel
  but this information was not directly used in determining
  $Q_{cor}$.}\@.

{\bf Uniformity Correction $U(i)$:} Cosmic ray muons that passed right
through the detector were used to correct for differences in the
response of the individual ends of every scintillator strip. This
calibration simultaneously removed many detector effects: scintillator
light output; gain of PMTs and electronics; different PMT quantum
efficiencies and pixel collection efficiencies; fibre differences;
optical connector transmission efficiencies; and many other
differences. Overall, the uniformity calibration aims to equalise the
response such that a hit at the centre of all strips gives an equal
response at every end of every strip. For hits away from the centre of
a strip, an additional correction for attenuation is required. For the
data presented herein, the uniformity calibration was done once and a
2\% statistical precision was achieved.

{\bf Attenuation Correction $A(i,x)$:} A correction for light
attenuation along the length of strips was made for this
analysis. This was necessary because the centre of the beam spot
struck the detector up to 2~cm away from the centre of the first plane
(depending on momentum and particle type). The correction was applied
to the positron shower or muon track as a whole because it was not
possible to reconstruct the position of individual hits at the
centimetre level. Four distinct groups of strip-ends had a correction
applied to them: horizontal/vertical strips and the two strip-ends of
each. A correction factor of $\pm$0.2\%/cm was applied to horizontal
and vertical strips depending on where the beam struck the face of the
detector relative to the centre. The corrections applied to the two
ends of a given strip were inversely correlated: if one end was
corrected up the other was corrected down by the same amount. The
correction factor of $\pm$0.2\%/cm was calculated using MC simulations
of positrons striking the detector in a wide range of different
positions, both horizontally and vertically. As described in
section~\ref{sec:detmc} the attenuation along the strips was
parameterised in the MC by fitting to data from both cosmic muons and
bench-top measurements using a radioactive source.

{\bf Signal Scale Correction $S$:} Stopping muons are used to set an
overall energy scale~\cite{ref:Jeff}. A muon energy unit~(MEU) is
defined, which corresponds to the energy deposited by an approximately
0.8~GeV muon passing perpendicularly through a 1~cm thick MINOS
scintillator plane (the average absolute energy deposition by such a
muon is approximately 2~MeV). An energy deposition of 1~MEU yields
about 4~photoelectrons~(PE) per strip-end. The same scale factor is
used for both readout systems; i.e. the MEU scale has no impact on the
relative response comparison. However, the MEU scale is defined to
allow calorimetric response comparisons among the three MINOS
detectors.


\section{Calibration Uncertainties}
\label{sec:calibsyst}

\subsection{Drift Correction}
\label{sec:calibsyst_drift}

The drift in the absolute response of the two readout systems with
time was of the order of several percent~\cite{ref:CalDet1} over
several weeks of running. This was mostly due to large temperature
variation in the experimental area of up to 10$^\circ$C\@. However,
the absolute temperature dependence of the two readout systems was
found to be almost identical and hence the error on the relative
response was a negligible 0.1\%. This was determined from 1~GeV/c test
beam positron data sets taken at regular intervals.

\subsection{Uniformity Correction}
\label{sec:calibsyst_uniformity}

\begin{figure}
  \centerline{\includegraphics[width=\linewidth]{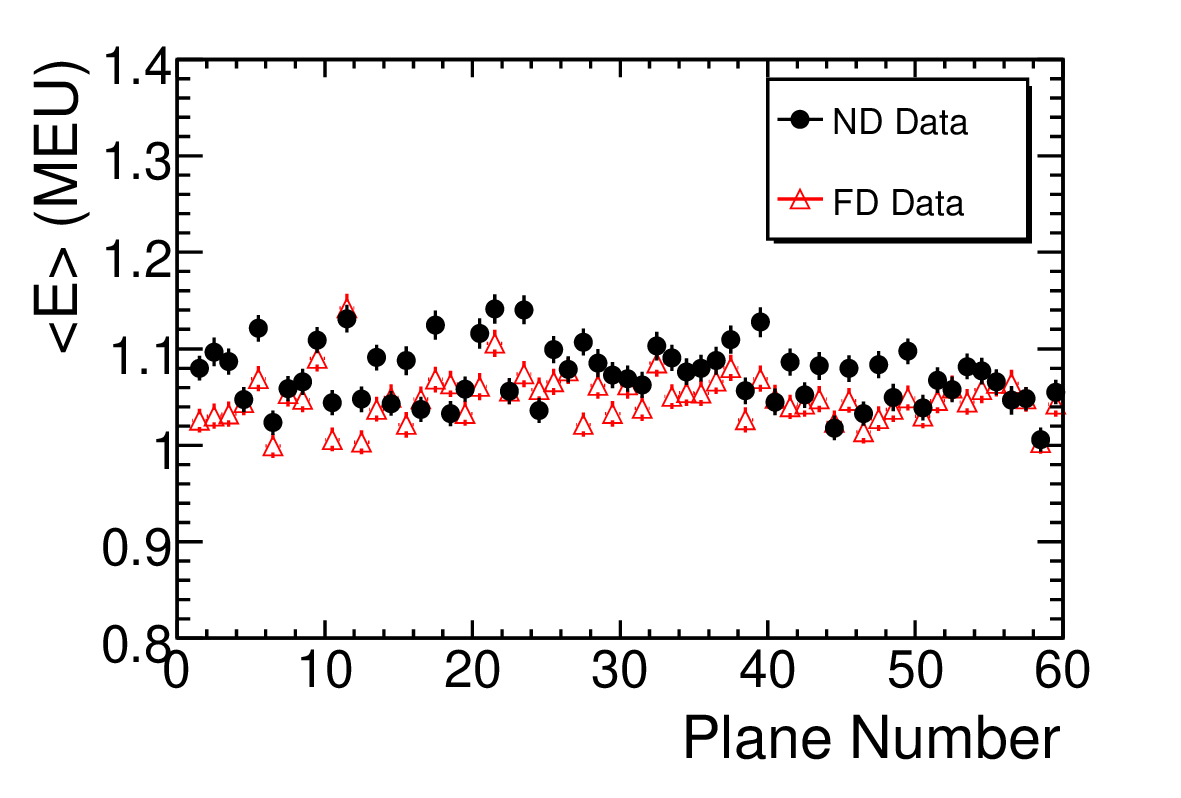}}
  \caption{Average calibrated energy deposited versus plane by 
    $>$2~GeV/c test beam muons. The error bars show the statistical
    error, thus the jitter from point to point is indicative of
    systematic error in the uniformity calibration. The observed
    response spread of the points is 2.4\% and 3.0\% for the FD and ND
    readout systems respectively.}
  \label{fig:S2SResponse}
\end{figure}

The performance of the uniformity correction is illustrated in
Fig.~\ref{fig:S2SResponse} by plotting the average calibrated response
per plane for test beam muons with momentum greater than
2~GeV/c. These muons travel through the detector by entering at the
centre of the first plane and exiting within a 25~cm radius of the
last plane's centre. The FD and ND readout systems have a 2.4\% and
3.0\% spread in response from plane to plane respectively, which
demonstrates the size of the systematic uncertainties on the
uniformity correction.

\subsection{Attenuation Correction}
The uncertainty on the attenuation correction arises primarily from
determination of the correction factor, which was calculated using
positron MC generated at a wide range of beam spot positions (as
described in Section~\ref{sec:calib}). A linear fit to the difference
in the response of the two readout systems as a function of beam spot
position was performed. The systematic spread of the residuals of
these high statistics points was 0.25\% and is taken as the error.

\subsection{Non-Linearity Measurements}
\label{sec:calibsyst_linearity}

The response of the readout system to light from the scintillator is not
perfectly linear due to non-linearities in the PMT and electronics
response. To incorporate the correct level of non-linearity into the MC
simulation a data driven approach was taken. The light-injection system
was used to measure the average non-linearity (1-measured/expected) of
all readout channels for the two systems separately. These measurements
were then used in the MC simulation to induce a non-linear response.

\begin{figure}
  \centerline{\includegraphics[width=\linewidth]{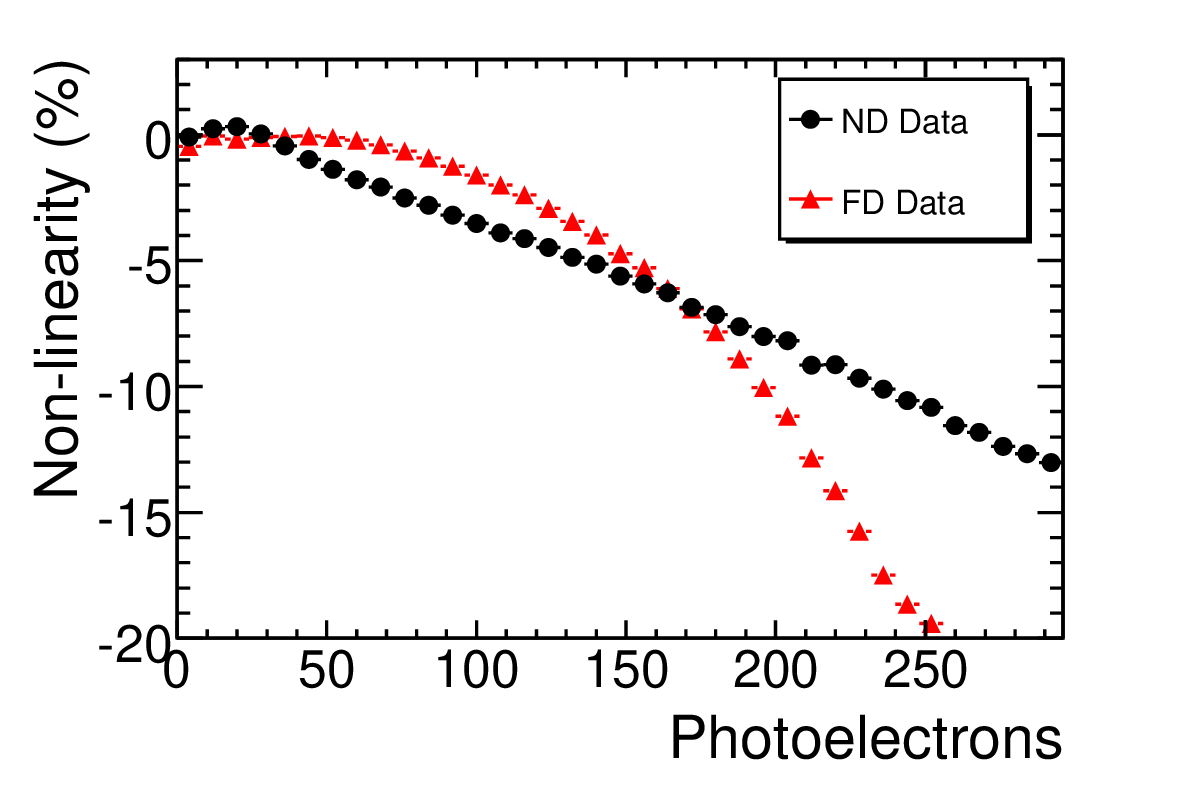}}
  \caption{Average non-linearity of the ND and FD readout systems 
    versus the number of PE injected by the light injection calibration
    system. The difference in the non-linearity between the two readout
    systems can be clearly seen.}
  \label{fig:NonLin}
\end{figure}

The data from the light-injection system are shown in
Fig.~\ref{fig:NonLin}, where the non-linearity of each readout system
is plotted as a function of the number of expected photoelectrons. PIN
photodiodes that measured the quantity of injected light with a high
degree of linearity were used to determine the expected number of
photoelectrons. The ND PMTs are more non-linear than the FD PMTs. A
fit to the CalDet data gave -0.05\%/PE and -0.03\%/PE deviations from
linearity for the ND and FD PMTs
respectively~\cite{ref:MyThesis}. However, as can be seen from
Fig.~\ref{fig:NonLin}, the PMT non-linearity does not explain all the
data. The shape of the overall (PMT plus electronics) readout system
non-linearity is quite different for the two cases. This occurs
because of a saturation effect in the FD electronics that becomes
significant for signals above about 100~PE\@. Beyond 200~PE the FD
readout system can be seen to become significantly more non-linear
than the ND system. Overall, the non-linearities of both readout
systems are the same to within 2\% up to about 200~PE\@.

No correction for non-linearity was applied to the data or the
MC\@. This approach was taken to investigate and understand the
relative effects of the non-linearity. Furthermore, by considering
this worst case scenario it is possible to set a maximum error on the
relative energy calibration between the two detectors. This maximum
error can then be reduced by performing a non-linearity calibration in
the Near and Far detectors. However, since the non-linearity of the
two readout systems is similar, the systematic error on the relative
energy scales is expected to be small for the range of energy
depositions relevant to MINOS\@.


\section{Strip Occupancy and PMT Crosstalk Comparison}
\label{sec:stripOccAndPmtXTalk}

\begin{figure}
  \centerline{\includegraphics[width=0.93\linewidth]{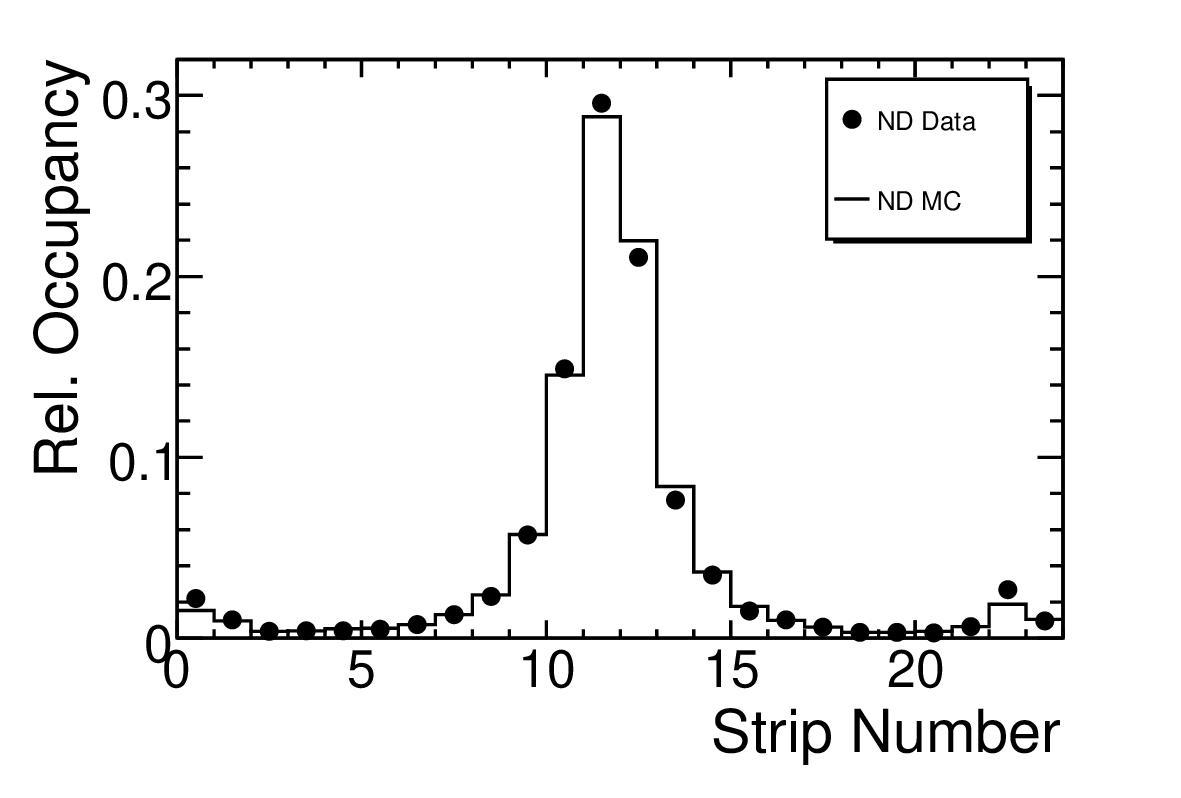}}
  \centerline{\includegraphics[width=0.93\linewidth]{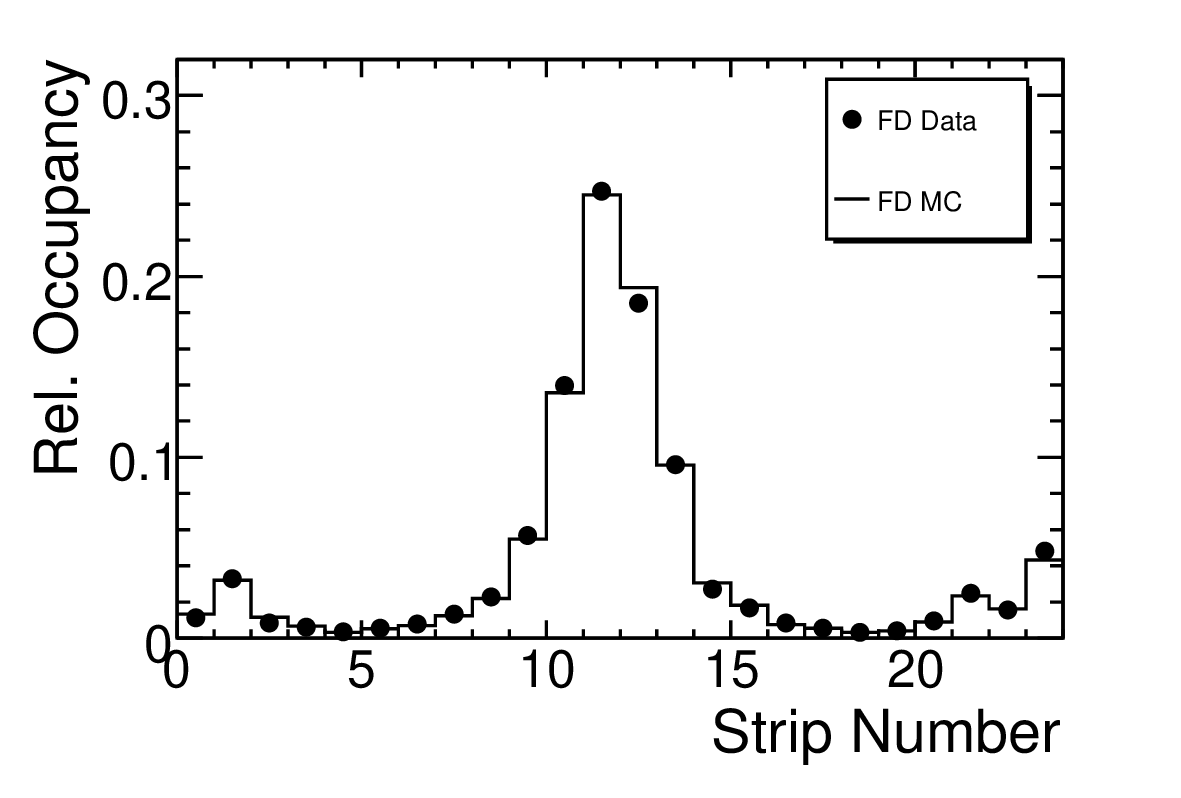}}
  \caption{Transverse hit profile of $>$2~GeV/c test beam muons as
    measured with the ND (top) and FD (bottom) readout systems for
    data and MC\@. The central peak is mostly formed from strips that
    the muon actually passed through. Whereas, in contrast, the peaks
    towards the edges of the detector (strip number 0\,-\,3 and
    20\,-\,23) are dominated by crosstalk hits. These data are from
    planes with strips that run horizontally.}
  \label{fig:transprof}
\end{figure}

The strips that register a response for a given particle interaction
is not always the same for the two readout systems. Differences in
strip occupancy\footnote{Strip occupancy is a count of the number of
  times a specific strip is hit, normalised by the total number of
  strips registering a hit.} can arise due to PMT crosstalk, the
average number of photons striking the PMTs and readout
thresholds. The effect of crosstalk on the strip occupancy can be
clearly demonstrated by using test beam muons since they only strike
the central few strips all along the length of the
detector. Fig.~\ref{fig:transprof} shows the relative strip occupancy
for test beam muons with momentum greater than 2~GeV/c. The central
peak is mostly formed from strips that the muon actually passed
through. In contrast, the peaks towards the edges of the detector
(strip number 0\,-\,3 and 20\,-\,23) are dominated by crosstalk
hits. The strip occupancy at the detector edges is different between
the two readout systems due to the different cabling configurations
and PMTs used. Overall, it can be seen that the strip occupancy is
well reproduced by the MC for both readout systems.

In addition to studying the strip occupancy, detailed studies were
made of the crosstalk at the PMT level. The crosstalk arises primarily
when photoelectrons injected on one pixel leak into the dynode chains
of neighbouring pixels. Thus, crosstalk can be quantified as
$Q_{Neighbour}/Q_{injected}$, where $Q_{injected}$ is the charge on
the anode of the pixel where light was injected, and $Q_{Neighbour}$
is the charge that appears on the anodes of the neighbouring
pixels. In-situ measurements of this crosstalk using muons in CalDet
gave values of $5.2\pm0.1\%$ and $4.8\pm0.1\%$ for the FD and ND
respectively~\cite{ref:MyThesis}. However, since 16 of the 64 pixels
on each ND PMT were not used, the observed crosstalk fraction
associated with strips dropped to $3.2\pm0.1\%$, consistent with
expectations\footnote{The naive expectation is for the crosstalk
  associated with strips to drop by 16/64 to 3.6\%. However, effects
  arise from exactly which 48 of the 64 pixels are connected to strips
  since some pixels receive more crosstalk than others.}.


\section{Calorimetric Response Comparison}
\label{sec:results}

The calorimetric response of the readout systems are compared in two
distinct ways: a ``hit-by-hit'' comparison and an ``event-by-event''
comparison. In the hit-by-hit comparison the difference in the
response from the two ends of individual strips is taken. For the
event-by-event comparison, it is the total sum of the response of all
the strip-ends read out by one system that is compared with the total
sum of the response of all the strip-ends read out by the other
system. In this paper, the differences are characterised by a relative
response asymmetry, $A_{\mathcal N/ \mathcal F}$. In general form,
$A_{\mathcal N/ \mathcal F}$ is written as
\begin{eqnarray}
\label{eq:anf}
  A_{\mathcal N/ \mathcal F} 
  & = & 
  \frac{\mathcal N-\mathcal F}{\frac{1}{2} (\mathcal N+\mathcal F)},
\nonumber
\end{eqnarray}
where $\mathcal N$ and $\mathcal F$ are the calorimetric response of
the ND and FD readout systems respectively. The data are presented
here as the average hit-by-hit asymmetry, $<~\!\!\!A_{\mathcal N/
  \mathcal F}^{Hit}\!\!\!~>$, and as the average event-by-event
asymmetry, $<~\!\!\!A_{\mathcal N/ \mathcal F}^{Event}\!\!\!~>$. It
should be noted that for a comparison to be made in the hit-by-hit
case, a response above threshold on both ends of the strip was
required. In contrast, strips with a response at just one end were
included in the event-by-event comparison. Depending on the particle
energy, approximately 90\,-\,95\% of the total event energy comes from
strips with a hit on both ends. The remainder comes from strips with a
response on only one end and that fraction is consistent with PMT
crosstalk, fluctuations in the production of photoelectrons, and
readout thresholds. Hits at very low energy, below a threshold of
0.14~MEU (approximately 0.5 photoelectrons) were removed from the
analysis. This cut was made on fully calibrated units to equalise the
threshold for the two readout systems and was applied to both data and
MC.

\subsection{Hit-by-Hit Response Comparison}

\begin{figure}
  \centerline{\includegraphics[width=\linewidth]{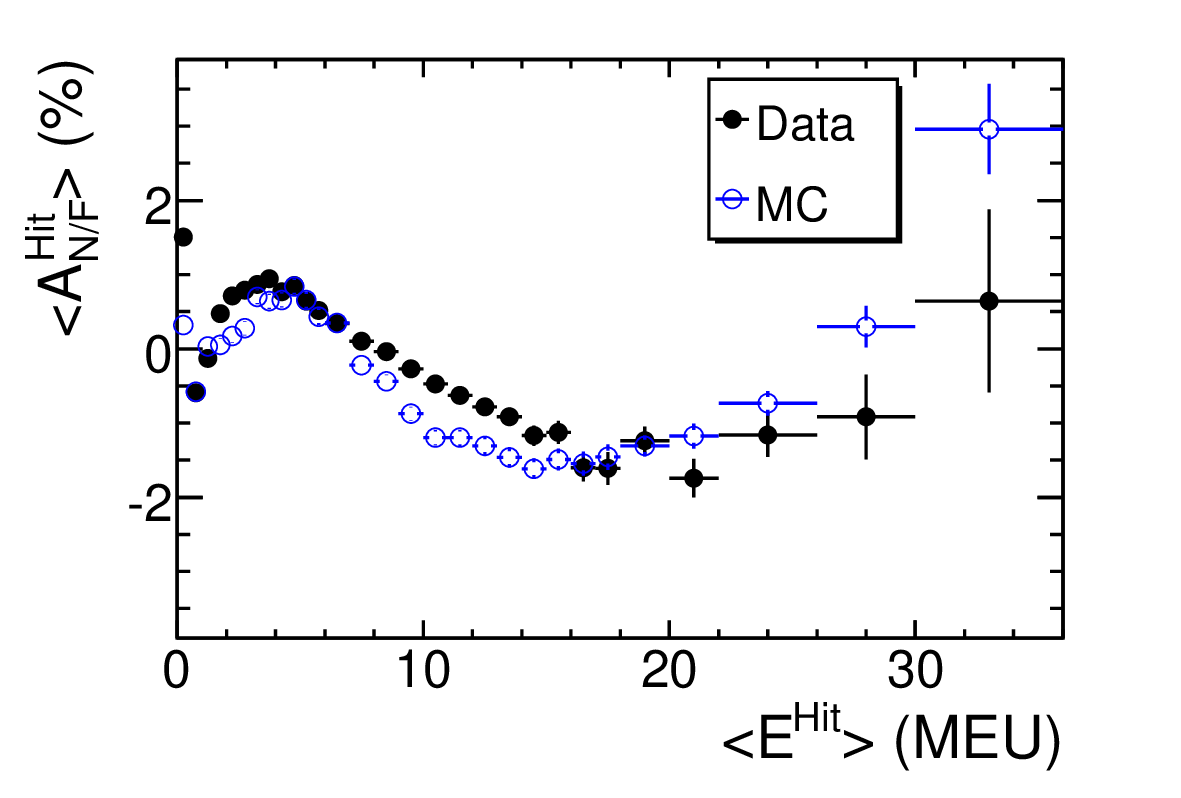}}
  \caption{Average hit-by-hit asymmetry ratio, $<~\!\!\!A_{\mathcal N/
      \mathcal F}^{Hit}\!\!\!~>$, as a function of the average energy
    deposition in the scintillator strip hit for data and MC\@. It can
    be seen that difference between the response of ND and FD readout
    systems changes by up to 2.5\% depending on the energy deposition.}
  \label{fig:AnfHit}
\end{figure}

The average hit-by-hit asymmetry, $<~\!\!\!A_{\mathcal N/ \mathcal
  F}^{Hit}\!\!\!~>$, is shown in Fig.~\ref{fig:AnfHit} as a function
of energy deposition. The data from positrons of between
0.6\,-\,6.0~GeV/c were used to cover the range of energy depositions
from 0\,-\,36~MEU\@. The uniformity correction forces the asymmetry to
be zero at around 1.5~MEU, which is the average energy deposition per
strip of the cosmic ray muons used for the calibration of
CalDet. Variation of $<~\!\!\!A_{\mathcal N/ \mathcal
  F}^{Hit}\!\!\!~>$ as a function of energy deposition is best
described by considering 2 distinct regions: above and below 4~MEU\@.

$<~\!\!\!A_{\mathcal N/ \mathcal F}^{Hit}\!\!\!~>$ changes by up to
2\% below 4~MEU with data and MC agreeing to 1\%. The general trend is
that $<~\!\!\!A_{\mathcal N/ \mathcal F}^{Hit}\!\!\!~>$ increases by
about 1.5\% between 0.5\,-\,4~MEUs of energy deposited, which means
that the ND response is increasing faster with energy deposited than
the FD response. This variation is caused by the difference in the way
the readout threshold is applied for the two readout systems: the ND
electronics applies a threshold to each 18.8~ns integration sample
separately, whereas the FD electronics only applies a threshold to the
total integration sample (see Section~\ref{sec:detdet} for more
details). As the energy deposited increases, the fraction below
readout threshold tends towards zero and so this effect is no longer
significant. This trend is well described by the MC simulation, which
shows a maximal deviation of less than 0.6\% from the data in all but
the lowest energy bin. The effect of the 0.14~MEU threshold for
including hits in the analysis was to reduce the asymmetry by 3\% in
the lowest energy bin but it had a negligible effect on the other
bins.

Between 4\,-\,15~MEU the slope goes the opposite way with
$<~\!\!\!A_{\mathcal N/ \mathcal F}^{Hit}\!\!\!~>$ decreasing with
energy deposited by 2.5\%. This is caused by the fact that the ND PMTs
are less linear than the FD PMTs, as shown in
Fig.~\ref{fig:NonLin}. For energy depositions between 18\,-\,36~MEU,
the trend is reversed with the slope of $<~\!\!\!A_{\mathcal N/
  \mathcal F}^{Hit}\!\!\!~>$ becoming positive again. The location of
this inflection is consistent with the light level at which the FD
readout system non-linearity starts to increase faster than that of
the ND (see Fig.~\ref{fig:NonLin}). The data and MC agreement in the
region between 4\,-\,36~MEU is better than 1\%. It is important to
note that the results in this region quantify the maximum error in the
case when a non-linearity correction was not applied. Whereas, when a
non-linearity correction is applied the differences between the two
readout systems at energy depositions above 4~MEU will be reduced.

In summary, the hit-by-hit data presented here show energy dependent
differences between the calorimetric response of the two readout
systems. These differences are shown to be caused by threshold effects
as well as PMT and electronics non-linearity. The fact that the
differences in the two readout systems are well simulated is
particularly important for the relative calibration between the MINOS
detectors. This result shows that as long as data and MC are made to
agree at the energy of the muons used for calibration (around 1.5~MEU)
then the MC can be used to correct for the relative differences between
the ND and FD readout systems.

\subsection{Event-by-Event Response Comparison}

\begin{figure}
  \centerline{\includegraphics[width=\linewidth]{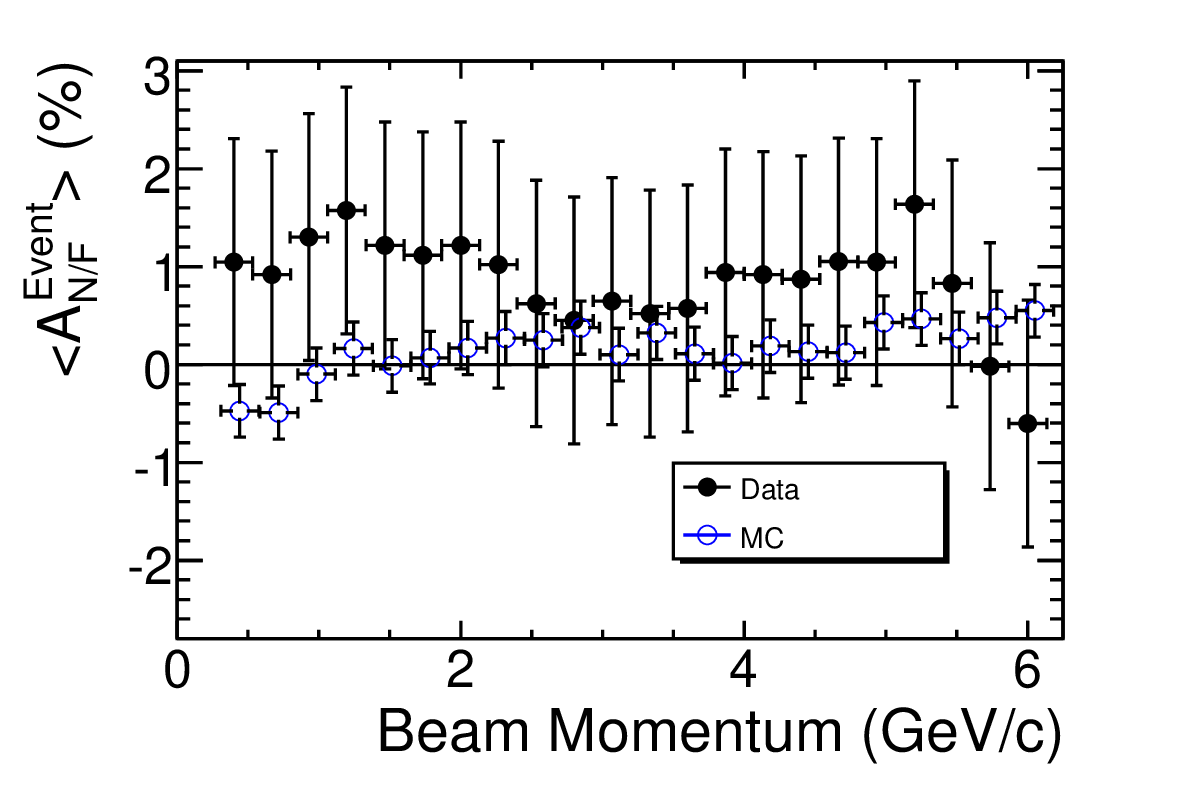}}
  \caption{Average event-by-event asymmetry ratio,
    $<~\!\!\!A_{\mathcal N/ \mathcal F}^{Event}\!\!\!~>$, as a function of the
    test beam momentum. The error bars show the total systematic error
    (statistical errors are negligible).}
  \label{fig:AnfEventMean}
\end{figure}

To make the comparison of the event-by-event response of the two
readout systems, data from test beam positrons between
0.6\,-\,6.0~GeV/c were used. Fig.~\ref{fig:AnfEventMean} shows the
average event-by-event asymmetry, $<~\!\!\!A_{\mathcal N/ \mathcal
  F}^{Event}\!\!\!~>$, as a function of the test beam momentum. It can be
seen that there is no significant energy dependence. This is an
important demonstration that the differences between the two readout
systems seen in the hit-by-hit comparison do not introduce an overall
{\bf relative} energy dependent bias between the ND and FD (despite
the absence of a non-linearity correction in the data presented
here). It should be noted that a correction of 2.0\% for the
difference in crosstalk between the two readout systems was made in
data and MC (see Section~\ref{sec:stripOccAndPmtXTalk}).

The average value of $<~\!\!\!A_{\mathcal N/ \mathcal F}^{Event}\!\!\!~>$
between 0.6\,-\,6.0~GeV/c is
\begin{eqnarray}
& & 1.2\pm1.3\% \;\; \mbox{for data, and}\nonumber\\
& & 0.3\pm0.3\% \;\; \mbox{for MC\@.}\nonumber
\end{eqnarray}
The statistical errors on both data and MC are negligible. The
systematic uncertainties arise from the following sources: the
uncertainty in the attenuation correction; the correction for
differing quantities of crosstalk associated with strips; and bias in
the uniformity correction.

The uncertainty on the attenuation correction gives a 0.25\% error on
$<~\!\!\!A_{\mathcal N/ \mathcal F}^{Event}\!\!\!~>$. The uncertainty
on the difference in the cross talk between the two readout systems
gives a 0.14\% error. For MC the systematic error from the uniformity
correction is by definition zero since the same calibration constants
are used to both induce differences between the response of all the
strip-ends in the simulation and then to calibrate out those
differences.

\begin{figure}
  \centerline{\includegraphics[width=\linewidth]{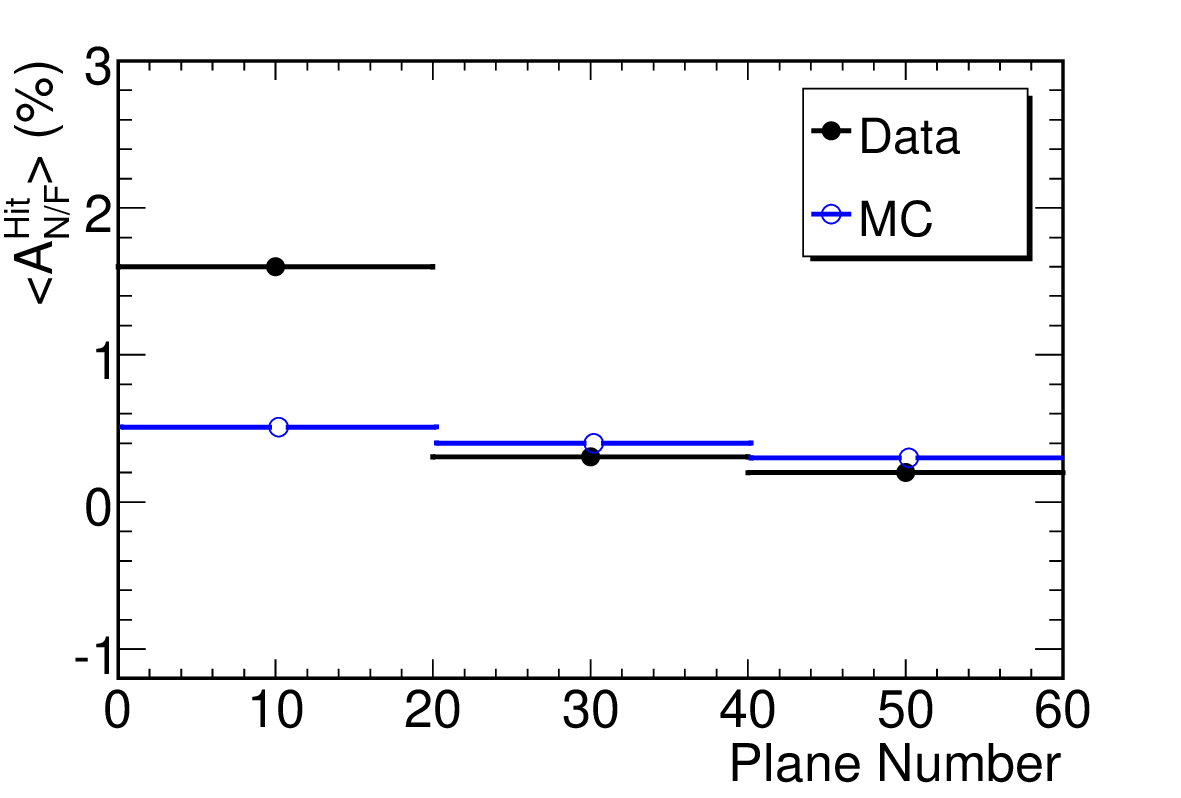}}
  \caption{$<~\!\!\!A_{\mathcal N/ \mathcal F}^{Hit}\!\!\!~>$ computed
    with test beam through going muons along the detector axis
    (statistical errors only). Data exhibits a difference in
    $<~\!\!\!A_{\mathcal N/ \mathcal F}^{Hit}\!\!\!~>$ between the
    first third and the rest of the detector, which is consistent with
    calibration systematic bias.}
  \label{fig:S2SError}
\end{figure}

To estimate the systematic uncertainty on $<~\!\!\!A_{\mathcal N/
  \mathcal F}^{Event}\!\!\!~>$ that arises from the uniformity
calibration, muons from the test beam with a momentum of more than
2~GeV/c were used. The value of $<~\!\!\!A_{\mathcal N/ \mathcal
  F}^{Hit}\!\!\!~>$ obtained using these muons is shown as a function
of position along the detector in Fig.~\ref{fig:S2SError}. The data
show a 1.1\% difference between the first third of the detector and
the back two thirds. In contrast, the asymmetry in the MC simulation
is close to uniform throughout as expected. The choice of dividing the
detector into thirds was a trade off. The muons predominantly sample
only one strip in each plane in the first 20 planes or so. In
contrast, the positrons shower and sample 3\,-\,5 strips per
plane. Sampling just a few strips at the front of the detector with
the muons would not be a fair estimate of the asymmetry. Thus, the
trade off was between increasing the number of strips sampled by the
muons and their relevance to positrons. Twenty planes was chosen since
it encompasses close to 100\% of the positrons energy, less than this
would have removed some of the strips that are hit by the higher
energy positrons.

The asymmetry measured using test beam muons is interpreted as the
error on the uniformity calibration by a process of elimination. The
known sources of error from the attenuation correction and crosstalk
are not large enough to cause the observed asymmetry, leaving the
uniformity correction as the only possible candidate. The measured
asymmetry value of 1.1\% is used as an estimate of the systematic
error on the uniformity calibration and is the dominant error on
$<~\!\!\!A_{\mathcal N/ \mathcal F}^{Event}\!\!\!~>$.

In summary, three sets of particles have been used to measure the
asymmetry in different ways. Cosmic muons were used for the uniformity
calibration with the aim of removing the inherent asymmetry that is
present in the uncalibrated responses of the two readout systems. Test
beam muons were then used as an independent sample of the same
particle type in assessing the residual error on the uniformity
calibration. Positrons from the test beam, with their large spread of
energy depositions, were used to assess the final level of asymmetry:
showing that it was both consistent with zero within the total
systematic uncertainty and independent of test beam momentum.

\subsection{Readout System Resolution}

\begin{figure}
  \centerline{\includegraphics[width=\linewidth]{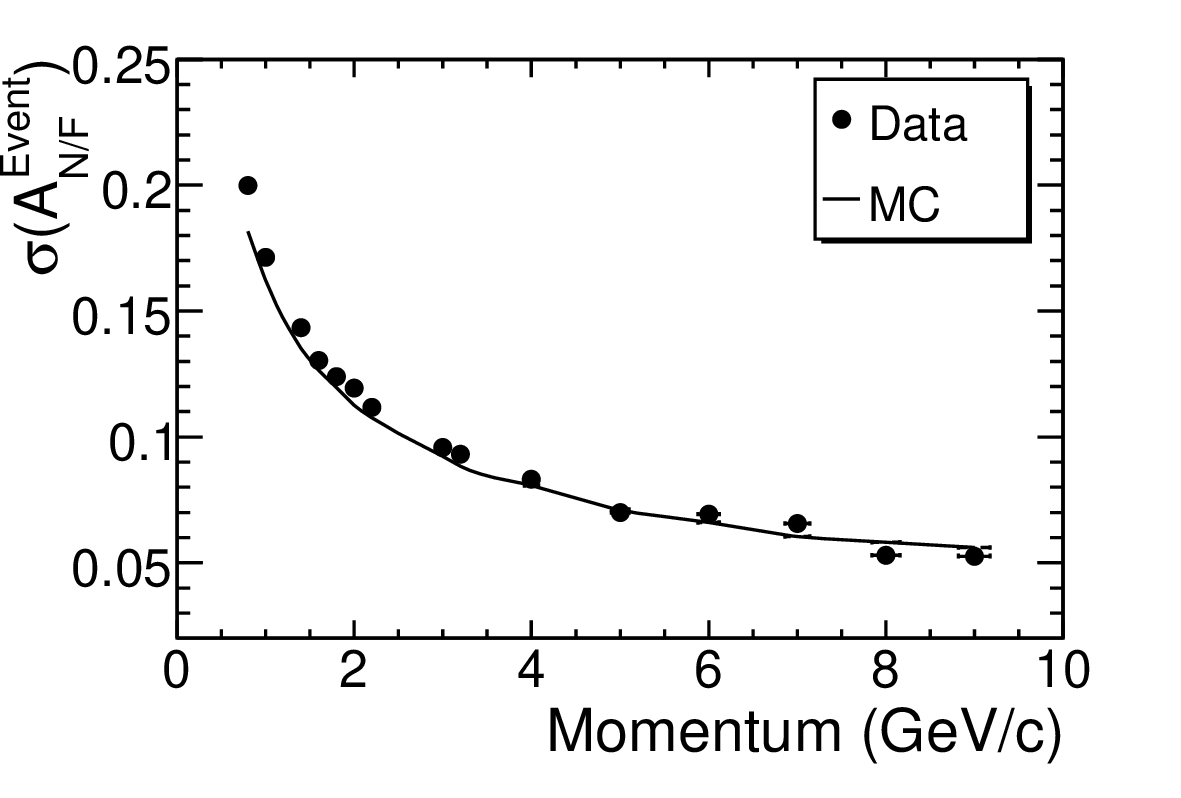}}
  \caption{The readout system resolution (width of 
    $A_{\mathcal N/ \mathcal F}^{Event}$) as a function of positron
    momentum. This figure demonstrates the size of the fluctuations that
    occur in the readout when the same physics event is observed by both
    Near and Far systems. It can be seen that the size of the
    fluctuations decreases with energy and is well reproduced by the MC
    simulation.}
  \label{fig:AnfEventWidth}
\end{figure}

The data described in this paper were obtained with the scintillator
strips instrumented at one end with the ND readout system and at the
other end by the FD system. The advantage of this approach was that it
made the comparison of the two systems insensitive to fluctuations in
the energy loss of the particles themselves and the simulation of such
fluctuations in the MC\@. The fluctuations observed in $A_{\mathcal N/
  \mathcal F}^{Event}$ from event to event are due to fluctuations in
the detector readout and the absorption/re-emission of photons in the
WLS fibre. The dominant source of fluctuations arises in the
conversion of photons to photoelectrons at the PMT face. For example,
a 1~GeV/c positron yields about 180~PE and leads to a 15\% width. The
width of the distribution of $A_{\mathcal N/ \mathcal F}^{Event}$ is
shown in Fig.~\ref{fig:AnfEventWidth} as a function of positron
momentum. It can be seen that the size of the fluctuations decreases
with energy and is well reproduced by the MC simulation.

\section{Conclusions}
\label{sec:conclusions}

The MINOS Calibration detector has acquired test beam data in a
configuration that allowed direct comparison of the two different
readout systems used for the Near and Far detectors. These data were
used to understand the systematic differences between the readout
systems and to demonstrate that the differences can be controlled to the
level required by MINOS for neutrino oscillation measurements.

Overall, comparison of the hit-by-hit response showed that the
calibration procedure reduced the differences to better than 2.5\%
over a wide range of energies, even without correction for
non-linearity of the PMTs and electronics. To understand the residual
differences there are two regions to consider: energy depositions
above and below 4~MEU\@. The relative hit-by-hit response of the two
readout systems below 4~MEU changes by up to 2\% with the MC
reproducing the trend to 1\%. The cause for this difference is
demonstrated to be dominated by threshold effects in the
electronics. This affect could potentially bias the relative energy
calibration in MINOS\@. One way to avoid this bias is to calibrate the
MC so that it agrees with the data in each detector separately. Above
about 4~MEU the hit-by-hit response of the two systems differs by up
to 2.5\% with the MC reproducing the trends to better than 1\%. This
2.5\% variation is caused by differences in the non-linearity of the
two types of PMT and electronics used for the Near and Far detectors.

The event-by-event comparison showed that the differences in the
overall calorimetric response of the two readout systems were
consistent with zero to an accuracy of 1.3\% and 0.3\% for data and MC
respectively.  Furthermore, no significant energy dependence was
observed in positron data sets taken with test beam momentum settings
between 0.6\,-\,6~GeV/c. These results clearly demonstrate the ability
of the calibration procedure to reduce the inherent differences in the
total calorimetric response of the two readout systems to a low level.

Differences in strip occupancy were shown to arise due to effects of
crosstalk between pixels on the two types of multi-anode
photomultiplier tubes used. This crosstalk, which is different for the
two readout systems was demonstrated to be well reproduced by the MC
simulation: both in the fraction of the total energy deposition
dispersed as crosstalk and in its distribution across the pixels.

In the MINOS Near and Far detectors it is neutrino interactions that are
reconstructed and that involves a multiplicity of particles in the
hadronic shower. However, it is the combination of the result presented
here for single positrons and the demonstrated accuracy of the MC
simulation that is most important. These two results combined give
confidence that any effects arising from the two different readout
systems will be small enough so that they do not significantly impact
the neutrino oscillation measurements made by MINOS\@.


\section{Acknowledgements}

This work was funded in part by the UK Particle Physics and Astronomy
Research Council (PPARC), the US Department of Energy (DoE) and the
European Union (EU). We would like to thank CERN for providing the
test beams and support. Special thanks are due to L. Durieu and
M. Hauschild for their help throughout this project. We are grateful
for the engineering support provided by T.~Durkin, M.~Proga, D.~Atree,
J.~Trevor, J.~Hanson, M.~Williams, P.~Groves and G.~Sillman and
electronics support from C.~Nelson, B.~Luebke, T.~Fitzpatrick.


\end{document}